\documentclass[11pt]{article}

\advance \textheight by \topskip
\makeatletter
\@addtoreset{equation}{section}
    
\makeatother

\def\vp{\vec{p}}
\newcommand{\p}{\partial}
\newcommand{\half}{{\scriptstyle{\frac{1}{2}}}}
\newcommand{\fourth}{{\scriptstyle{\frac{1}{4}}}}

\usepackage{amsmath,amssymb}

\begin{document}

\title{
{\bf   Anyon wave equations and the noncommutative plane
}}

\author
{{\sf Peter A. Horv\'athy}\thanks{
Permanent address: {\it Laboratoire de Math\'ematiques et de
Physique Th\'eorique}, Universit\'e de Tours (France).
E-mail: horvathy@univ-tours.fr.}
{\sf\ and Mikhail S. Plyushchay}\thanks{
Also: {\it Institute for High Energy Physics},
Protvino (Russia).
E-mail: mplyushc@lauca.usach.cl}
\\[4pt]
{\small \it
Departamento de F\'{\i}sica,
Universidad de Santiago de Chile}\\
{\small \it Casilla 307, Santiago 2, Chile}\\
}
\date{}

\maketitle

\begin{abstract}
     The ``Jackiw-Nair'' non-relativistic limit
     of the relativistic anyon equations provides us with
     infinite-component wave equations of the
     Dirac-Majorana-L\'evy-Leblond type
     for the ``exotic'' particle, associated with the
     two-fold central extension of the planar Galilei group.
     An infinite dimensional representation
     of the Galilei group is found. The velocity operator is
     studied, and the observable coordinates
     describing a noncommutative plane are identified.
\end{abstract}

\vskip.5cm\noindent

\section{Introduction}

What kind of wave equation do anyons satisfy?
The question has haunted researchers since the beginning
\cite{JNany, Plany, MD, RHeis}.
A related problem is to find a wave equation for  ``exotic''
particles \cite{LSZ,HP1,DH}, associated with the two-fold
central extension of the planar Galilei group \cite{exotic}.
(Remember that exotic Galilean symmetry implies the
noncommutativity
of the position coordinates \cite{LSZ,DH} and, at the
classical level, can be deduced from spinning anyons as a
tricky non-relativistic  limit \cite{JaNa}).

Thirty years ago \cite{Dirac} Dirac  proposed a new
wave equation for particles with internal structure.
Adapting his ideas to the plane, one of us (M.P.)
has derived a $(2+1)$-dimensional version of Dirac's new
equation that describes
relativistic anyons and usual fields of integer and
half-integer spin  in a unified way \cite{RHeis}.
This theory, briefly summarized  in Section \ref{Major}
below,
will be the starting point for our new developments here.
Using the Fock space representation,
we consider two kinds of non-relativistic limits.
Both limits yield infinite sets of first-order equations.
In the first type the spin is kept fixed and, at each
step,
the new component gets a factor of  $c^{-1}$. The first two
components yield the non-relativistic ``Dirac
equation"
put forward by L\'evy-Leblond in the sixties 
\cite{LL}.
The ``Jackiw-Nair (JN) limit'' \cite{JaNa}
yields instead a genuinely
infinite component system, (\ref{infLL}),
of ``L\'evy-Leblond-type'' equations.
These latter are invariant with respect to the
two-fold centrally extended exotic Galilei group, whose
action
can be derived from the Poincar\'e-symmetry of the anyon.
Presented in terms of the initial
commuting coordinates, our system
undergoes a non-relativistic Zitterbewegung.
Identifying the observable, Zitterbewegung-free
quantities results
in Galilei-covariant coordinates, which describe a
noncommutative plane.

\section{Wave equations for relativistic
anyons}\label{Major}

Let us briefly summarize the theory of anyons we start with.
It was emphasised by Jackiw and Nair \cite{JNany}
(see also \cite{Plany}) that
anyons, just like usual fields of integer and half-integer
spin,
correspond to irreducible representations of the planar
Poincar\'e group \cite{PPrep},
labeled with two Casimir invariants, namely
\begin{eqnarray}
     (p_{\mu}p^{\mu}+m^2c^2)|\Psi\rangle=0,\label{KG}
     \\[5pt]
     (p_{\mu}J^\mu-smc)|\Psi\rangle=0,\label{PaLu}
\end{eqnarray}
where $J_\mu$ is the ``spin part'' of the total angular
momentum operator
\begin{equation}
{\cal J}_\mu=-\epsilon_{\mu\nu\lambda}x^\nu p^\lambda
+J_\mu.
\label{calJ}
\end{equation}
The ${\cal J}_\mu$, together with $p_\mu$, generate
the (2+1)D Poincar\'e group, via the commutation
relations
\begin{equation}
[x_\mu,p_\nu]=i\eta_{\mu\nu},\quad
[x_\mu,x_\nu]=[p_\mu,p_\nu]=[J_\mu,x_\nu]=[J_\mu,p_\nu]=0,
\quad
[J_\mu,J_\nu]=-i\epsilon_{\mu\nu\lambda}J^\lambda,
\label{COMMUT}
\end{equation}
where $\eta_{\mu\nu}=diag (-1,+1,+1)$,
$\epsilon^{012}=1$.
The choice
$J_\mu=-\frac{1}{2}\gamma_\mu$ and $s=\pm1/2$
reduces Eq. (\ref{PaLu}) to the Dirac equation.
Similarly, the choice
$(J_\mu)^\sigma{}_\rho=-i\epsilon^\sigma{}_{\mu\rho}$
for $s=\pm1$
leads to the topologically massive vector field equation
\cite{JNany}.
In these two cases (only), the quadratic
equation (\ref{KG}) follows from the linear equation
(\ref{PaLu}).

To describe anyons for which the spin parameter $s$ can take
any real value, we have to resort to the
infinite-dimensional
half-bounded unitary representations of the planar Lorentz
group  of the discrete type series
$D^+_\alpha$ (or, $D^-_\alpha$) \cite{JNany,Plany,RHeis}.
For these representations, characterized by
$J_\mu J^\mu=-\alpha(\alpha-1)$
and
$J_0=diag (\alpha +n)$ ($-J_0=diag (\alpha+n)$),
$\alpha>0$, $n=0,1,\ldots,$ equation (\ref{PaLu})
reduces to a (2+1)-dimensional analog
of the Majorana equation \cite{Majorana}. The pair of
equations (\ref{KG}), (\ref{PaLu}) describes an anyon field
with spin $s=\alpha$ (or, $s=-\alpha$ for the choice of
$D^-_\alpha$).
Here, as for usual fields with $|s|=j>1$,
the Klein-Gordon equation (\ref{KG})
is independent of the Majorana equation.

The wave functions chosen by
Jackiw and Nair \cite{JNany} (see also \cite{MD})
carry a reducible
representation of the planar Lorentz group.
Their wave equation, (4.10), needs
therefore be
supplemented by further conditions, their eqns. (4.12a) and
(4.12b).

Our primary aim here is to derive a non-relativistic
model from the relativistic anyon. This could be attempted
 in the framework of \cite{JNany}; subsidary
conditions would lead to complications, though.

Another line of attack is provided by
the \lq\lq new wave equation''  proposed by Dirac in 1971
\cite{Dirac}, which
describes a particle in (3+1)D, endowed with an internal
structure.
 Dirac's idea has been applied to anyons in $2+1$ dimensions
\cite{RHeis}~:
one starts with the one-dimensional deformed Heisenberg
algebra
\begin{equation}
     \big[a^-,a^+\big]=1+\nu R,
     \qquad
     R^2=1,
     \qquad
     \{a^{\pm},R\}=0,
     \label{defHeis}
\end{equation}
where $\nu$ is a real (deformation) parameter.
  $N=\frac{1}{2}\{a^+,a^-\}-\frac{1}{2}(\nu+1)$
  plays the role of a number operator,
  $[N,a^\pm]=\pm a^\pm$,
  allowing us to present a reflection operator $R$
  in terms of $a^\pm$: $R=(-1)^N=\cos\pi N$.
For any $\nu>-1$, this algebra admits
an infinite-dimensional unitary representation realized on
Fock space\footnote{For negative odd integer values
$\nu=-(2k+1)$, $k=1,2,\ldots$, one gets finite, namely
$(2k+1)$-dimensional representations.
Then the equations (\ref{aanyoneqn}) describe a spin-$j$
field with $j=k/2$, which has states with both signs of the
energy
\cite{RHeis}. From now on, we only consider the
infinite-dimensional case.}. The vacuum state
is distinguished by the relation $a^-|0\rangle=0$, and
$|n\rangle$, $N|n\rangle=n|n\rangle$,
is given by $|n\rangle=C_{n}(a^+)^n|0\rangle$,
where $C_{n}$ is a normalization
coefficient.
Fock space is decomposed into even and odd subspaces
defined
by $R|\psi\rangle_{\pm}=\pm|\psi\rangle_{\pm}$, which
correspond to $n$ even or odd.
The quadratic operators
\begin{equation}
     J_{0}=\frac{1}{4}\{a^+,a^-\},
     \qquad
     J_{\pm}\equiv J_1\pm iJ_2=\frac{1}{2}(a^{\pm})^2
\label{a2}
\end{equation}
generate the so$(1,2)$ Lorentz algebra,
see the last relation in (\ref{COMMUT}).
The so$(1,2)$ Casimir is $J_{\mu}J^{\mu}=-s(s-1)$
with $s=\fourth(1\pm\nu)$ on the even/odd subspaces.
The quadratic operators (\ref{a2}), together with the
linear operators
$$
     L_{1}=\frac{1}{\sqrt{2}}(a^++a^-),
	\qquad
     L_{2}=\frac{i}{\sqrt{2}}(a^+-a^-),
$$
extend the Lorentz algebra into an osp$(1|2)$ superalgebra:
$\{L_\alpha,L_\beta\}=4i (J\gamma)_{\alpha\beta}$
$[J_{\mu},L_{\alpha}]=
\half(\gamma_{\mu})_{\alpha}^{\ \beta}L_{\beta}$,
where the two-by-two $\gamma$-matrices are in the
Majorana representation,
$(\gamma^0)_{\alpha}^{\ \beta}=
-(\sigma^2)_{\alpha}^{\ \beta}$,
$(\gamma^1)_{\alpha}^{\ \beta}=
i(\sigma^1)_{\alpha}^{\ \beta}$,
$(\gamma^2)_{\alpha}^{\ \beta}=
i(\sigma^3)_{\alpha}^{\ \beta}$.
The antisymmetric tensor
$\epsilon_{\alpha\beta}$, $\epsilon_{12}=1$,
provides us with a metric for the spinor indices,
$(\gamma^\mu)_{\alpha\beta}=(\gamma^\mu)_\alpha{}
^\rho\epsilon_{\rho\beta}$.
The space-time coordinates $x^\mu$ and
conjugate momenta $p_\mu$ are independent from the
internal variables, $[x^\mu,a^\pm]=[p_\mu,a^\pm]=0$;
hence, due to (\ref{calJ}),
the $L_{\alpha}$ form a (2+1)D spinor.
With all these ingredients at hand, now we posit our
linear-in-$p_\mu$ anyon equations \cite{RHeis}
\begin{equation}
     Q_{\alpha}|\psi\rangle=0,
     \qquad\hbox{where}\qquad
     Q_{\alpha}=\left(R
     \big(p_{\mu}\gamma^\mu\big)_{\alpha}^{\ \beta}
     +mc\epsilon_{\alpha}^{\ \beta}\right)L_{\beta}.
     \label{aanyoneqn}
\end{equation}
\goodbreak

Since the $Q_\alpha$ and the total angular momentum operator
given by Eqs.  (\ref{calJ}) and (\ref{a2}) satisfy
the relation $[{\cal J}_\mu,Q_\alpha]=\half
(\gamma_{\mu})_{\alpha}^{\ \beta}Q_{\beta}$,
  (\ref{aanyoneqn}) is a covariant {\it spinor} set
of equations. (In contrast, the corresponding operators
in (\ref{KG}) and (\ref{PaLu}) are so$(1,2)$ scalars).
In our Dirac-type approach the consistency
conditions
\begin{equation}
\left\{Q_{\alpha},Q_{\beta}\right\}|\psi\rangle=0
\qquad\hbox{and}\qquad
\left[Q_{\alpha},Q_{\beta}\right]|\psi\rangle=0,
\label{conscond}
\end{equation}
restricted to the even subspace,
imply equations (\ref{KG}) and (\ref{PaLu}),
whereas they force the odd part
to vanish identically, $|\psi\rangle_-=0$ \cite{RHeis}.
We shall restrict our considerations
to even states henceforth. Then our wave functions
carry an irreducible representation of
the planar Lorentz group and do not necessitate therefore
any subsidary condition. 
Compared to the wave functions of
\cite{JNany} (or, of \cite{MD}) which carry 
additional vector (spinor)
indices,
ours are more economical, as they boil down to
a scalar-like field.

An insight is gained when we
write (\ref{aanyoneqn}) on Fock space.
Expanding as $|\psi\rangle=\sum_{n\geq 0}\psi_{n}|2n\rangle$
yields, for each $n\geq0$, a pair of coupled equations which
only involve neighbouring components, namely
\begin{equation}
         \left\{\begin{array}{ll}
	\sqrt{n+2s}\,(mc+p_0)\psi_{n}
	-\sqrt{n+1}\,p_+\psi_{n+1}=0,
	\\[12pt]
	\sqrt{n+2s}\,p_-\psi_{n}
	+\sqrt{n+1}\,(mc-p_0)\psi_{n+1}=0,
     \end{array}\right.
\label{anyoneqn}
\end{equation}
where $p_\pm\equiv p_1\pm ip_2$.
The second equation in (\ref{anyoneqn}) yields the
recurrence relation
\begin{equation}
\psi_{n+1}=-\sqrt{\frac{n+2s}{n+1}}\,\frac{p_-}{mc-p_0}\,
\psi_n.
\label{n+1}
\end{equation}
Substitution into the first equation
reproduces, for each $\psi_{n}$, the Klein-Gordon equation
(\ref{KG})~: $(p^2+m^2c^2)\psi_n=0$.
To see that our anyon equations
imply also Eq. (\ref{PaLu}) which fixes the second (spin)
Casimir invariant, let us first observe that
the spin generators in (\ref{a2}) act on Fock space
according to
\begin{equation}
     \begin{array}{lll}
\langle 2n|J_0|\psi\rangle=(s+n)\psi_n,
\\[10pt]
\langle 2n|J_+|\psi\rangle=\sqrt{n(n-1+2s)}\,\psi_{n-1},
\\[10pt]
\langle 2n|J_-|\psi\rangle=\sqrt{(n+1)(n+2s)}\,\psi_{n+1}.
\end{array}
\label{j}
\end{equation}
Then
 taking into account Eqs. (\ref{j})
yields the matrix form of (\ref{PaLu}), namely
\begin{equation}
     \langle 2n|p_{\mu}J^\mu-smc|\psi\rangle=0.
\end{equation}
Equations (\ref{anyoneqn}) describe therefore
an infinite-component field of mass $m$, whose
(anyonic) spin, $s=\fourth~(1~+~\nu~)~>~0$,
is fixed by the value of the deformation parameter
$\nu>-1$. In the rest frame system $\vec{p}=0$
we find also that, just like for the usual
Majorana equation \cite{Majorana}, the sign of the energy of
those states described by  (\ref{anyoneqn}) is
necessarily positive \cite{RHeis}, $p^0>0$.
Note also that the positive energy anyonic states with
negative spin values, $s=-\fourth(1+\nu)<0$, can be
obtained by changing
the sign before the mass term $mc$ in (\ref{aanyoneqn})
and positing the so(1,2) generators
$J_0=-\fourth \{a^+,a^-\},\, J_{\pm}=-\half(a^{\mp})^2$.
Then repeated application of (\ref{n+1}),
together with equation (\ref {PaLu}) projected
onto the vacuum state, provides us with the
momentum-representation solution to the anyon equations
(\ref{anyoneqn}),
\begin{equation}
    \begin{array}{ll}
     \psi_{n}(p)=(-1)^n\sqrt{\frac{2s\cdot(2s+1)\cdots
     (2s+n-1)}{n!}}
     \displaystyle\left(\frac{p_{1}-ip_{2}}{mc-p_{0}}\right)
     ^n\psi_{0}(p),
     \\[8pt]
\psi_0(p)=\delta(p^0-\sqrt{m^2c^2+\vec{p}\,{}^2}\,)\,
\psi(\vec{p}).
\end{array}
\label{psi0}
\end{equation}

\section{``L\'evy-Leblond'' equations for exotic particles}

Now we consider the non-relativistic limit of our anyon
equations. For the purpose we note that, consistently
with Eqs. (\ref{psi0}),
 all components vanish in the rest frame, with the exception
of $\psi_0$, which has energy $p^0=mc$.
Then, putting $p_0=-ic^{-1}\p_t$ and
$$
|\psi\rangle=e^{-imc^2t}|\phi\rangle=
e^{-imc^2t} \sum_{n\geq 0}\phi_{n}|2n\rangle,
$$
the Fock-space equations (\ref{anyoneqn}) yield the
first-order system
$$
   \begin{array}{ll}
	\displaystyle \sqrt{n+2s}\,ic^{-1}\p_t\,\phi_n
	+\sqrt{n+1}\,p_+\,\phi_{n+1}=0,
	\\[10pt]
	\sqrt{n+2s}\,p_-\,\phi_{n}+
	\sqrt{n+1}\, (2mc-ic^{-1}\p_t)\,\phi_{n+1}=0.
   \end{array}
$$

We can now consider {\it two types} of non-relativistic
limits. Firstly, let us keep the spin $s$ fixed and let
$c\rightarrow \infty$.
In this limit the subsequent components get always
multiplied by
$c^{-1}$, $\phi_{n+1}\sim c^{-1}\phi_{n}$.
Only keeping terms up to order $c^{-2}$, we are left with
just the
first two  equations. Calling $\phi_{0}=\Phi$ and
$c\phi_{1}=\chi$, they read
\begin{equation}
     \left\{\begin{array}{ll}
     \displaystyle
     i\p_{t}\Phi+\displaystyle\frac{1}{\sqrt{2s}}\,p_+
     \,\chi=0,
     \\[12pt]
   \displaystyle\frac{1}{2m}\,p_-\,\Phi+
   \frac{1}{\sqrt{2s}}\,\chi=0.
	\label{LL}
     \end{array}\right.
\end{equation}
These equations form already a closed system, which
generalizes from $s=1/2$ to any $s>0$
the two-component ``non-relativistic ``Dirac'' equation
introduced by L\'evy-Leblond \cite{LL}.
Let us emphasise that in this ``ordinary'' non-relativistic
limit, relativistic
spin simply becomes non-relativistic spin, still denoted by
$s$.
It should be remembered, however, that (\ref{LL})
represents only
the two first leading equations in the expansion in
the small parameter ${|\vec{p}|}/{mc}$, and that the NR
limit of the anyon system of (any) spin $s$
 contains, unlike for the non-relativistic
limit of the Dirac equation, an infinite number of
components $\phi_n\sim ({|\vec{p}|}/{mc})^n\phi_0$.
Redefining the higher-order components as
$\Phi_{n}=c^n\phi_{n}$
yields indeed
\begin{equation}
     \left\{\begin{array}{ll}
     \displaystyle
     i\p_{t}\Phi_{n}+\sqrt{\displaystyle\frac{n+1}{n+2s}}\,
     p_{+}\,\Phi_{n+1}=0,
     \\[12pt]
   \displaystyle\frac{1}{2m}\, p_{-}\,\Phi_{n}
   +\sqrt{\displaystyle\frac{n+1}{n+2s}}\,\Phi_{n+1}=0.
	\label{genLL}
     \end{array}\right.
\end{equation}

Rather then pursuing these investigations, we
 focus our attention to another, more
subtle limit, considered by Jackiw and Nair \cite{JaNa}.
As spin in $(2+1)$D is a continuous parameter, we can indeed
let it diverge so that $s/c^2$ tends to a
finite limit:
\begin{equation}
c\rightarrow \infty,\qquad
s\rightarrow \infty,\qquad
\frac{s}{c^2}=\kappa.
\label{JNlim}
\end{equation}
Then all components remain of the
same order, and we get an infinite number of equations
\begin{equation}
     \left\{\begin{array}{ll}
     i\p_{t}\phi_{n}+\displaystyle\sqrt{\frac{n+1}{2
     	\kappa}}\,
	p_{+}\,\phi_{n+1}=0,
	\\[16pt]
	\displaystyle\frac{1}{2m}\,p_{-}\,\phi_{n}
	+\sqrt{\frac{n+1}{2\kappa}}\,\phi_{n+1}=0.
     \end{array}\right.
     \label{infLL}
\end{equation}
These are the first-order, infinite-component
Dirac-Majorana-L\'evy-Leblond  type equations
we propose to describe our free ``exotic'' system.

Eliminating one component shows, furthermore, that each
component
satisfies the free Schr\"odinger equation
$
i\p_{t}\phi_{n}=\big({\vec{p}\,{}^2}/{2m}\big)\phi_{n}
$
(cf. (\ref{KG})).
For further discussion we observe that, grouping
all ``upper'' and all ``lower equations''
collectively, (\ref{infLL}) can also be presented as
\begin{equation}
D|\phi\rangle=0,\qquad
\Lambda|\phi\rangle=0.
\label{DL}
\end{equation}
The  second equation here can be viewed as a
quantum constraint which specifies the physical subspace.
It allows us to express all components in terms of the first
one:
\begin{equation}
\phi_n=(-1)^n\left(\frac{\kappa}{2}\right)^{\frac{n}{2}}
\frac{1}{
\sqrt{n!}}
\, \left(\frac{p_1-ip_2}{m}\right)^n\phi_0,
\end{equation}
cf. (\ref{psi0}).
\goodbreak

\section{Exotic Galilei symmetry}\label{Galsym}

Both types of our non-relativistic systems considered above
are indeed invariant under Galilean transformations. The
generators of Galilei boosts
can be derived  from the relativistic
Lorentz generators as the $c\to\infty$ limits of
\begin{equation}
     {\cal K}_i=-\frac{1}{c}\,\epsilon_{ij}{\cal J}_j.
     \label{LorGalgen}
\end{equation}
Put
$
\delta \phi_n=i\delta b_j\, \langle 2n|{\cal K}_j|\phi
\rangle,
$
where $\delta b_j$ are the transformation parameters.
Dropping terms of $o(c^{-2})$ for the  (spin-$s$)
L\'evy-Leblond system (\ref{LL})
we get, using (\ref{calJ}) and (\ref{j}),
\begin{equation}
     \begin{array}{ccc}
\delta \Phi=&\delta b_i\Big(mx_{i}-tp_{i}\Big)\Phi,&
\\[6pt]
\delta \chi=&\delta b_i\Big(mx_{i}-tp_{i}\big)\chi
&-\displaystyle\sqrt{\frac{s}{2}}\,(\delta b_1+i\delta b_2)
\, \Phi.
\end{array}
\label{sLLrep}
\end{equation}
For spin $s=1/2$, we recover the formula of L\'evy-Leblond
\cite{LL}.
This representation is conventional in that the
boosts commute, $[{\cal K}_{i},{\cal K}_{j}]=0$.

Let us stress, however, that (\ref{LL})  is just an
$o(c^{-2})$ truncation of an infinite-component system
(\ref{genLL}), whose
Galilean symmetry could be established by recursion.
Let us indeed posit that Galilei boosts act on the first
component $\Phi_{0}$ as on a scalar, i. e. as on $\Phi$ in
(\ref{sLLrep}).
Then the action on the second component $\Phi_{1}$ can be
deduced seeking it in the form
$\delta\Phi_{1}=\alpha_{0}\Phi_{0}+\alpha_{1}\Phi_{1}$
and requiring the field equations to be satisfied. The
procedure
could be continued for the next component, etc. Explicit
formulae that generalize (\ref{sLLrep}) are
not illuminating as they soon become rather complicated.

Let us now identify the Galilean symmetry of the
anyon field system (\ref{infLL}), obtained by
the JN limit (\ref{JNlim}).
The  Galilean boost generators (\ref{LorGalgen}),
acting on infinite-dimensional Fock space,
can be found from equations (\ref{calJ}) and (\ref{j}) as
\begin{equation}
\begin{array}{cccc}
\langle 2n|{\cal K}_1|\phi\rangle=
&(mx_1-tp_1)\phi_n
&-&i
\sqrt{\frac{\kappa}{2}}\,(\sqrt{n+1}\,
\phi_{n+1}-\sqrt{n}\,\phi_{n-1}),
\\[10pt]
\langle 2n|{\cal K}_2|\phi\rangle=
&(mx_2-tp_2)\phi_n
&+
&\sqrt{\frac{\kappa}{2}}\,(\sqrt{n+1}\,
\phi_{n+1}+\sqrt{n}\,\phi_{n-1}).
\end{array}
\label{Gbo}
\end{equation}
The first, diagonal-in-$\phi_{n}$
terms here represent the ordinary Galilean
symmetry (as for a scalar particle). The terms which mix
the  components $\phi_{n\pm1}$
are associated with the ``internal" structure
of the system.
The remarkable feature of the boosts (\ref{Gbo}) is that,
owing precisely to the internal structure, they satisfy
\begin{equation}
[{\cal K}_1,{\cal K}_2]=-i\kappa
\label{exoboostcommrel}
\end{equation}
rather then commute.
Relation (\ref{exoboostcommrel}) is the
hallmark of ``exotic'' Galilean symmetry \cite{LSZ,
DH,exotic},
further discussed in the next  Section.

To identify the generator of rotations of the system,
we note that the JN limit of the relativistic
angular momentum ${\cal J}_0$ (just like of the energy)
diverges \cite{DHcontr}.
Omitting the divergent part
and taking into account Eq. (\ref{calJ})
and the first equation from (\ref{j}) yields
\begin{equation}
\langle 2n|{\cal J}|\phi\rangle\equiv
\langle 2n|{\cal J}_0|\phi\rangle_{renorm}= (\epsilon_{ij}
x_ip_j +n)
\phi_n.
\label{J+n}
\end{equation}
The internal contribution $n\phi_n$
is essential for
establishing the correct commutation relations
$[{\cal J},{\cal K}_j]= i\epsilon_{jk}{\cal K}_k$.

\section{The velocity operator}

The Hamiltonian of the system is
the generator of time translations. It is
convenient to introduce the non-relativistic
counterparts of the relativistic bosonic operators
we denote (for reasons which will become
clear later)  by $v_\pm$. They act
on  ``non-relativistic'' Fock space spanned by
$|n\rangle_v=|2n\rangle$, $n=0,1,\ldots$,
\begin{equation}
     v_+|n\rangle_v=- (2/\kappa)^{1/2}
     \sqrt{n+1}\,|n+1\rangle_v,
\qquad
     v_-|n\rangle_v=- (2/\kappa)^{1/2}
     \sqrt{n}\,|n-1\rangle_v.
     \label{ve}
\end{equation}
The factor $-(2/\kappa)^{1/2}$
is included for later convenience.

The operators $v_\pm$ here are in fact the JN limits
of the translation invariant part
of relativistic Lorentz generators,
$-(c/s)J_{\pm}\to v_{\pm}$,
cf. (\ref{LorGalgen}).
The JN limit of $\langle 2n|J_+|\psi\rangle$
yields, e.g., by (\ref{j}),
${}_v\langle
n|(-c/s)J_{+}|n-1\rangle_v\to-(2/\kappa)^{1/2}\sqrt{n}$,
i.e.  (\ref{ve}).
These ``internal'' operators span an (undeformed) Heisenberg
algebra
$[v_-,v_+]=2\kappa^{-1}$, or,
putting $v_\pm=v_1\pm iv_2$,
\begin{equation}
[v_j,v_k]=-i\kappa^{-1}\epsilon_{jk}.
\label{vi}
\end{equation}
As it follows from the relativistic relations
(\ref{COMMUT}), the $v_i$
commute with the ``external'' canonical coordinates
$x_i$ and momenta $p_i$,
$ 
[x_i,p_j]=i\delta_{ij},
\;
[x_i,x_j]=[p_i,p_j]=[v_i,x_j]=[v_i,p_j]=0.
$ 

Any state of our system can now be decomposed over Fock
space, $|\phi\rangle=\sum_{n\geq 0}\phi_n|n\rangle_v$,
where the field components $\phi_n$ are taken either in
coordinate ($x_i$), or in momentum ($p_i$) representation.
Writing the field equations (\ref{infLL})  in the form (\ref
{DL}) with
$
D=i\p_t-\frac{1}{2}\,p_+v_-,
$
$
\Lambda=v_--m^{-1}\, p_-,
$
one would be tempted to view $\frac{1}{2}\,p_+v_-$
as a Hamiltonian. This is, however, not correct, as
$\frac{1}{2}\,p_+v_-$ is not Hermitian.
Our clue is that  (\ref{DL}) is equivalent
to  a set of equations of the same form, but with $D$
changed into
$D-\frac{1}{2}m v_+ \Lambda$.
The system of field equations
(\ref{infLL}) can finally be represented in the equivalent
form
\begin{equation}
     \left\{\begin{array}{c}
D|\phi\rangle=0
\\[10pt]
\Lambda|\phi\rangle=0
\end{array}\right.,
\qquad
\begin{array}{cc}
D=i\p_t-{\cal H},\hfill
&{\cal H}=\vec{p}\cdot\vec{v}-\frac{1}{2}m\, v_+v_-,
\\[10pt]
\Lambda=v_--\frac{1}{m}\,p_-.\quad&
\end{array}
\label{tilDL}
\end{equation}
The Hermitian operator ${\cal H}$ here can be identified
with the Hamiltonian of the system.
On the physical subspace defined by
$\Lambda|\phi\rangle_{phys}=0$, our
${\cal H}$ reduces to the free expression,
\begin{equation}
     {\cal H}|\phi\rangle_{phys}=H|\phi\rangle_{phys},
     \qquad H=\frac{\vec{p}\,{}^2}{2m}.
     \label{LSZQM}
\end{equation}
Thus, we recover the  framework proposed in
\cite{LSZ} on grounds of canonical quantization.
The {\it quadratic} expression in (\ref{LSZQM})
 comes from our eliminating $\phi_{n+1}$
using (\ref{ve}).
Note that, consistently with (\ref{tilDL}),
the physical states are just the
coherent states of the Heisenberg algebra corresponding to
the operators $v_\pm$,
$$
v_-|\phi\rangle_{phys}= m^{-1}\,p_-
|\phi\rangle_{phys},\qquad
|\phi\rangle_{phys}
\propto e^{-\frac{1}{2}\sqrt{\theta}\,p_-v_+}|0\rangle
_v,
$$
where we have introduced the notation \cite{DH}
\begin{equation}
\theta=\frac{\kappa}{m^2}.
\label{theta}
\end{equation}

The generators (\ref{Gbo}),
(\ref{J+n}) can be presented in the form
\begin{equation}
{\cal K}_i=mx_i-tp_i+\kappa\epsilon_{ij}v_j,\qquad
{\cal J}=\epsilon_{ij}x_ip_j+\frac{1}{2}\kappa v_+v_-.
\label{KJv}
\end{equation}
Let us note for further reference that
\begin{equation}
{\cal H}=\kappa^{-1}(\epsilon_{ij}{\cal K}_ip_j
-m{\cal J}).
\label{HGalilei}
\end{equation}

In the representation (\ref{tilDL}),
(\ref{KJv}), we obtain the nontrivial commutation relations
of the two-fold centrally extended Galilei group
\cite{exotic}, namely
\begin{equation}
     \begin{array}{ccc}
&[{\cal K}_i,p_j]=im\delta_{ij},\qquad
[{\cal K}_i,{\cal K}_j]=-i\kappa\epsilon_{ij},&
\\[12pt]
&
[{\cal K}_i,{\cal H}]=ip_i,\qquad
[{\cal J},p_i]=i\epsilon_{ij}p_j,\qquad
[{\cal J},{\cal K}_i]=i\epsilon_{ij}{\cal K}_j.&
\end{array}
\label{Galg}
\end{equation}

Conversely, the exotic relations
(\ref{Galg}) fix the additional terms in
(\ref{KJv}). Seeking indeed the generators of Galilei
boosts and rotations in the from
${\cal K}_{i}=mx_{i}-tp_{i}+\Gamma_{i}$ and
${\cal J}=\epsilon_{ij}x_{i}p_{j}+\Sigma$, respectively,
where $\Gamma_{i}$ and $\Sigma$ commute with the
external operators $x_{i}$ and $p_{i}$, the Galilei
relations
in (\ref{Galg}) imply that necessarily
$[\Gamma_{i},\Gamma_{j}]
=-i\kappa\epsilon_{ij}$, i. e., the $\Gamma_{i}$ span a
Heisenberg
algebra. Similarly, these relations also fix $\Sigma$ as
$\Sigma=\frac{1}{2\kappa}\Gamma_{i}\Gamma_{i}+s_{0}$ where
$s_{0}$ is a constant. Calling
$\Gamma_{i}=\kappa\epsilon_{ij}v_{j}$
and putting $s_{0}=0$ results in (\ref{KJv}).
Our previous
(infinite dimensional) formulae can be recovered
by representing the $v_{\pm}$
operators on Fock space according to (\ref{ve}).

Using the non-relativistic commutation relations given
above,
the Hamiltonian ${\cal H}$ is seen to generate
the Heisenberg equations of motion
\begin{equation}
\frac{d x_i}{dt}= v_i,\qquad
\frac{dp_i}{dt}=0,\qquad
\frac{dv_i}{dt}=\frac{m}{\kappa}\,\epsilon_{ij}\,(v_j-
m^{-1}p_j),
\label{zit}
\end{equation}
which is the quantum counterpart of the classical
equations studied in \cite{LSZ,HP1}, and
can  be integrated at once to give $p_i(t)=const.$ and,
\begin{equation}
x_i(t)=m^{-1}\left(p_i t-\kappa  \epsilon_{ij}V_j(t)\right)
+const.,\qquad
V_\pm(t)=\exp(\mp im\kappa^{-1} t)\,V_\pm (0),
\label{evol}
\end{equation}
where
$
V_\pm=V_1\pm iV_2,
$
and
\begin{equation}
V_i=v_i-m^{-1}p_i.
\label{Vi}
\end{equation}
The noncommuting operators $v_i$
describe therefore the velocity of the system and,
as for a Dirac particle,
the coordinates perform a ``non-relativistic
Zitterbewegung".
The internal variable $V_{i}$ measures the extent the
momentum, $p_{i}$, differs from [$m$-times] the velocity.
In its terms, the operator $\Lambda$ in (\ref{tilDL})
which defines the physical-state constraint is
$
    \Lambda=V_-.
$

\section{Observable coordinates in the noncommutative plane}

According to Eq. (\ref{evol}), the time evolution of
the initial (commuting) coordinates $x_i$
is different from that of a usual
free non-relativistic particle.
This happens because the $x_i$ do not commute with the
operator $\Lambda$ that singles out the physical subspace,
$
[x_1,\Lambda]=-im^{-1},
$
$
[x_2,\Lambda]=-m^{-1}.
$
As a result, the position operators do not leave the
physical subspace invariant,
$\Lambda x_i|\phi\rangle_{phys}\neq 0$.
Hence, following Dirac, they cannot be viewed as observable
operators.
Like for a Dirac particle \cite{Davyd},
one can identify the observable coordinates
as those that do commute with $\Lambda$.
To find them, let us consider a
unitary transformation 
 generated by the operator
$
U=\exp \left(i\kappa m^{-1}\, \epsilon_{jk}p_jv_k
\right),
$
\begin{equation}
Uv_iU^{-1}=V_i,\qquad
Ux_iU^{-1}=X_i,
\qquad
Up_iU^{-1}=p_i,
\label{Unitary}
\end{equation}
where
\begin{equation}
X_i={\cal X}_i+\frac{\theta}{2}\epsilon_{ij}p_j,\qquad
{\cal X}_i=x_i+\kappa m^{-1}\,\epsilon_{ij}V_j.
\label{XVUni}
\end{equation}
Here  the constant $\theta$
is given by (\ref{theta}).
Since $[X_i,V_j]=[x_i,v_j]=0$ by construction, the
$X_i$
can be viewed as observable coordinate operators.
In terms of the new operators, the non-transformed
Hamiltonian  (\ref{tilDL}) reads
\begin{equation}
{\cal H}=\frac{\vec{p}\,{}^2}{2m}-\frac{m}{2}V_+ V_-,
\label{HamA}
\end{equation}
and we conclude that, consistently with  (\ref{evol}), the
$X_i$
evolve as coordinate operators  of a free non-relativistic
particle, namely as $X_i(t)=X_i(0)+tm^{-1}p_i$.
The dynamics of $V_\pm$ is in turn that of a harmonic
oscillator.
Note, however, that $V_+$ is not observable
($[V_+,\Lambda]=-2\kappa^{-1}\neq 0$),
whereas $V_-|\phi\rangle_{phys}=0$, cf. \cite{LSZ}.
Thus, the physical states are the vacuum states of the
harmonic oscillator-like internal operators $V_\pm$.

At last, the
${\cal X}_i=X_i-\frac{\theta}{2}\epsilon_{ij}p_j$
commute with the $V_j$ [and hence with $\Lambda$],
and are therefore also observable.
Due to the conservation of $p_i$,
they have the same evolution law as the $X_i$.
However, the coordinates ${\cal X}_i$, $i=1,2$, unlike
$X_i$, do not commute between themselves,
\begin{equation}
[{\cal X}_j,{\cal X}_k]=i\theta\epsilon_{jk}.
\label{XX}
\end{equation}
Their noncommutativity stems from
the noncommutativity of the velocity operators
and the related Zitterbewegung.
In terms of the operators (\ref{XVUni}),
(\ref{Vi}),
\begin{equation}
{\cal K}_i=m{\cal X}_i-tp_i+m\theta\epsilon_{ij}p_j,\qquad
{\cal J}=\epsilon_{ij}{\cal X}_ip_j+\frac{1}{2}\theta p_i^2
+\frac{1}{2}\kappa V_+V_-
\label{KJnew}
\end{equation}
cf. \cite{LSZ,HP1}.
The operators $X_i$,
${\cal X}_i$ and $V_i$ are 2D vectors by construction.
Unlike the initial $v_i$, the $V_i$
are invariant under Galilei boosts,
$[{\cal K}_i,V_j]=0$,
whereas for $X_i$ and ${\cal X}_i$ we get
$$ 
[{\cal K}_j,{\cal X}_k]=it\delta_{jk},
\qquad
[{\cal K}_j,X_j]
=it\delta_{jk}-i\frac{1}{2}m\theta\epsilon_{ij}.
$$ 
This means that the (observable)
${\cal X}_i$ transform  under Galilei boosts as
planar coordinate operators
and, consistently with (\ref{XX}), they describe a
noncommutative plane.
(In contrast, the operators $X_i$ {\it are not
Galilei-covariant}).
Note that the ${\cal X}_i$ and $X_i$ are analogous to the
Foldy-Wouthuysen and Newton-Wigner coordinates
for the Dirac particle, respectively, \cite{NW,Davyd}.

We conclude that our system described
by the equations (\ref{infLL})
represents a free, massive, non-relativistic field on the
noncommutative plane. Its reduction to the physical subspace
$V_-|\phi\rangle_{phys}=0$ yields the free exotic
particle introduced in \cite{DH}.
The latter is described by the noncommutative coordinates
${\cal X}_i$ and momenta $p_i$ (see eq. (\ref{XX}) and
$[{\cal X}_i,p_j]=i\delta_{ij}$, $[p_i,p_j]=0$),
whose dynamics is given by the usual quadratic Hamiltonian
$H=\vp\,{}^2/2m$;
the generators of the Galilei boosts
are given by eq.
(\ref{KJnew}), while the angular momentum operator is
${\cal J}=\epsilon_{ij}{\cal X}_ip_j+
\frac{1}{2}\theta\vp\,{}^2$ cf. \cite{DH}.

\section{The JN limit of the Majorana equation}

In the anyon context, the Majorana equation
appeared as the equation describing the quantum theory
of the higher derivative (2+1)D model of a relativistic
particle with torsion \cite{Plany,Ptorsion}, whose
Euclidean version emerged originally in relation to
Fermi-Bose transmutation  mechanism \cite{Polyakov}.
Like the original (3+1)D Majorana equation \cite{Majorana},
its (2+1)D analog admits three types of solutions,
namely massive ($p^2<0$ with spectrum $M_n=ms/S_n$,
where $S_n=s+n$ is spin of the corresponding state,
$n=0,1,...$),
massless ($p^2=0$) and tachyonic ($p^2>0$) ones.
Then requiring also the Klein-Gordon equation
eliminates the massless and tachyonic
sectors and singles out
the only  massive state with spin $S_0=s$ and mass $m$.
We focus therefore our attention to the massive sector,
and inquire about the JN limit of the Majorana equation.

In Fock space associated with the deformed Heisenberg
algebra, the matrix form of the Majorana equation,
$
\langle 2n|p_{\mu}J^\mu-smc|\psi \rangle=0,
$
reads
\begin{equation}
-[p_0 (s+n)+smc]\psi_n+\frac{1}{2}
[\sqrt{n(n-1+2s)}\,p_-\psi_{n-1}
+
\sqrt{(n+1)(n+2s)}\, p_+\psi_{n+1}]=0.
\label{majoreq}
\end{equation}
Separating, as before,
the divergent part of energy by putting
$\psi_n=e^{-imc^2t}\phi_n$,  the JN limit yields
\begin{equation}
i\p_t\phi_n=-\frac{m}{\kappa}\,n\,\phi_n -\frac{1}{\sqrt{2
\kappa}}
\big(\sqrt{n}\,p_-\phi_{n-1}+\sqrt{n+1}\,p_+\phi_{n+1}\big).
\label{FockMajor}
\end{equation}
Equation (\ref{FockMajor}) is nothing else
as the component form of the first of equations
(\ref{tilDL})~: its equivalent presentation is
\begin{equation}
{}_{v}\langle n|i\p_t -{\cal H}|\phi\rangle=0,
\label{MajHam}
\end{equation}
with  ${\cal H}$ the Hamiltonian operator in eq.
(\ref{tilDL}).
The spectrum  can be extracted at once from
the equivalent form  (\ref{HamA}) of the Hamiltonian.
In the representation where both the momentum operator
$\vec{p}$
and $V_+V_-$ are diagonal,
we find that the energy of the state
$|\vec{p},n\rangle$, $n=0,1,\ldots$, is
\begin{equation}
E_n(\vec{p})=\frac{1}{2m}\vec{p}\,{}^2-\frac{m}{\kappa}\,n.
\label{Epl}
\end{equation}
Taking into account the form of the angular momentum
operator (\ref{KJnew}), the quantum number $n$ can be
interpreted as internal angular momentum.
In other words, the JN limit of the Majorana equation
is  a kind of non-relativistic rotator, whose
internal angular momentum can take only nonnegative integer
values, and whose spectrum, (\ref{Epl}), is unbounded from
below.

Therefore, the only but crucial difference between our ``
exotic''
system  (\ref{infLL}) and the JN limit
of the (2+1)D Majorana equation (\ref{FockMajor})
is that the first system also requires the additional
equation $V_-|\phi\rangle=0$
(whose  component form is  the second equation in
(\ref{infLL})).
Like Gupta-Bleuler quantization
of the electromagnetic field, this additional condition
``freezes" the internal ``rotator" degree of freedom,
which is responsible for the negative contribution
to the energy.

To conclude this section, we note that our
results here are consistent with those obtained
in our previous paper \cite{HP1}, where, on the one hand,
we demonstrated that
the JN limit of the relativistic
particle with torsion yields,  at the classical level,
the acceleration-dependent  model of
Lukierski-Stichel-Zakrzewski
\cite{LSZ}, and, on the other hand,
showed that the quantum version of the latter
can be described in terms of our
noncommutative coordinates ${\cal X}_i$,
momenta $p_i$, and noncommuting internal `rotator'
variables $V_i$ [denoted in \cite{HP1}
by $Q_i$ ]. The
dynamics and transformation properties
with respect to exotic Galilean boosts and rotations
are given by (\ref{HamA}), (\ref{KJnew}).

\section{Discussion}

The origin of the noncommutative plane
can be traced back to the noncommutativity of velocities
 obtained as the  JN limit of  anyons, (\ref{aanyoneqn}).
The structure of the associated  Fock space as well as that
of the deformed Heisenberg algebra of
the initial relativistic anyon system
are rooted in the hidden Fock space structure of
the Majorana equation which underlies anyon theory.
The (2+1)D Majorana equation is based
on the half-bounded infinite-dimensional
unitary representations of the (2+1)-dimensional
Lorentz group.

In close analogy with the relativistic case,
(\ref{calJ}),
the structure of the two-fold extended
Galilei group fixes the additional ``exotic'' terms in the
conserved
quantities so that they generate the Heisenberg algebra and
provides us with a
noncommuting velocity operator.

The `lower' set of equations  in (\ref{infLL})
singles out the physical subspace of the system
on which, according to the second equation from
(\ref{tilDL}), the non-relativistic Zitterbewegung
disappears, $\langle  |v_i-m^{-1}p_i|
\rangle=0$, $|\rangle=|\phi\rangle_{phys}$.
On the physical subspace
the system is hence described by the Galilei-covariant
coordinates ${\cal X}_i$ of the noncommutative plane
and by the momenta $p_i$.

Our free system (\ref{infLL}) is equivalent
to the exotic particle model in \cite{DH}.
The two systems {\it are not} equivalent, however,
if we switch on an interaction:  the nonphysical
polarizations
$|\vec{p},n\rangle$,
$
({\kappa}/{2})\,
V_+V_-|\vec{p},n\rangle=n|\vec{p},n\rangle,
$
($n=1,2\ldots$),
could reveal themselves as  virtual states.

>From the viewpoint of representation theory,
the two-fold centrally extended planar Galilei group
has two Casimir operators, namely
\begin{equation}
{\cal C}_1=\epsilon_{ij}{\cal K}_ip_j-m{\cal
J}-\kappa{\cal H},\qquad
{\cal C}_2=p_i^2-2m{\cal H}.
\label{casimirs}
\end{equation}
By eq. (\ref{HGalilei}), the first Casimir is fixed
as a strong operator relation, ${\cal C}_1=0$.
On the other hand, due to eq.
(\ref{HamA}), we have
${\cal C}_2=m^2V_+V_-$.
Hence, the physical subspace constraint
$V_-|\phi\rangle=0$ fixes the second Casimir to vanish,
${\cal C}_2|\phi\rangle=0$
(whereas in the model of Ref. \cite{DH}
${\cal C}_2=0$ is a strong operator relation).
An (up to a constant factor) equivalent form of the second
 Casimir is
$$
\tilde{\cal C}_2={\cal J}-m^{-1}\epsilon_{ij}{\cal
 K}_ip_j- \frac{1}{2}\theta p_i^2.
 $$
Since in the rest frame system $p_i=0$
 it reduces to ${\cal J}$ (that, in turn,
 becomes proportional to $V_+V_-$),
 our second equation requires the internal spin to vanish.
These two conditions play a role analogous to
that of  (\ref{KG}), (\ref{PaLu}) for relativistic
anyons.

In contrast, for the JN limit of the Majorana
equation,
only the first Casimir operator is fixed,
${\cal C}_1=0$, and the second
 is left free. This results
in the appearance of negative-energy states
in the model of Ref. \cite{LSZ}.


\vskip 0.4cm\noindent
{\bf Acknowledgements}.
PAH is indebted to the {\it Departamento de F\'{\i}sica} of
{\it Universidad de Santiago de Chile} for hospitality
extended to him and also to P. Stichel for enlightening
discussions.
This work was partially supported by the
Grants 1010073 and 7010073 from FONDECYT (Chile).



\begin{thebibliography}{99}


\bibitem{JNany}
R. Jackiw and V. P. Nair,
{\it Relativistic wave equation for anyons}.
{\sl Phys. Rev.} {\bf D43} (1991) 1933.

\bibitem{Plany}
M. S. Plyushchay,
{\it Relativistic particle with torsion, Majorana equation
and fractional spin}.
{\sl Phys. Lett.} {\bf B262} (1991) 71;
{\it The model of relativistic particle with torsion}.
{\sl Nucl. Phys}. {\bf B 362} (1991) 54.

\bibitem{MD}
M. S. Plyushchay,
{\it Fractional spin: Majorana-Dirac field}.
{\sl Phys. Lett.} {\bf B273} (1991) 250;
{\it The model of a free relativistic particle with
fractional spin}.
{\sl Int. Journ. Mod. Phys}. {\bf A7} (1992) 7045.

\bibitem{RHeis}
M. S. Plyushchay,
{\it Deformed Heisenberg algebra and fractional spin field
in $(2+1)$ dimensions}.
{\sl Phys. Lett}. {\bf B320} (1994) 91
[\texttt{hep-th/9309148}];
{\it Deformed Heisenberg algebra, fractional spin fields and
supersymmetry without fermions}.
{\sl Ann. Phys.} {\bf 245} (1996) 339
[\texttt{hep-th/9601116}];
{\it Deformed Heisenberg algebra with reflection}.
{\sl Nucl. Phys.} {\bf B491} [PM] (1997) 619
[\texttt{hep-th/9701091}];
{\it R-deformed Heisenberg algebra, anyons and
$D=2+1$ supersymmetry}.
{\sl Mod. Phys. Lett.} {\bf A12} (1997) 1153
[\texttt{hep-th/9705034}].

\bibitem{LSZ}
J.~Lukierski, P.~C.~Stichel, W.~J.~Zakrzewski,
    {\it Galilean-invariant $(2+1)$-dimensional models with
    a Chern-Simons-like term and $d=2$ noncommutative
    geometry}.
    {\sl Annals of Physics (N. Y.)} {\bf 260} (1997) 224
    [\texttt{hep-th/9612017}].

\bibitem{HP1}
P.~A.~Horv\'athy and M.~S.~Plyushchay,
{\it Non-relativistic anyons, exotic Galilean symmetry and
noncommutative plane}.
{\sl JHEP} {\bf 0206} (2002) 033
[\texttt{hep-th/0201228}].
P. A. Horv\'athy,
{\it Mathisson's electron: noncommutative
mechanics \& exotic Galilean symmetry, 66 years ago}.
{\sl Acta Physica Polonica} {\bf 34}  (2003) 2611
[\texttt{hep-th/0303099}].

\bibitem{DH}
C.~ Duval and P.~A.~Horv\'athy,
    {\it The exotic Galilei group and the ``Peierls
    substitution''}.
{\sl Phys. Lett.} {\bf B 479} (2000) 284
[\texttt{hep-th/0002233}];
{\it Exotic Galilean symmetry in the non-commutative plane,
and the Hall  effect}.
{\sl Journ. Phys.} {\bf A 34} (2001) 10097
[\texttt{hep-th/0106089}].

\bibitem{exotic}
J.-M.~L\'evy-Leblond,
{\it Galilei group and Galilean invariance}.
in {\it Group Theory and Applications} (Loebl Ed.),
{\bf II}, Acad. Press, New York, p. 222 (1972);
 A. Ballesteros, N. Gadella and M.~del Olmo,
{\it Moyal quantization of 2+1 dimensional Galilean
systems}. {\sl Journ. Math. Phys.} {\bf 33} (1992) 3379;
 D.~R.~Grigore,
{\it The projective unitary irreducible representations
  of the Galilei group in $1+2$ dimensions}.
{\sl Journ. Math. Phys.} {\bf 37} (1996) 460
[\texttt{hep-th/9312048}];
  Y.~Brihaye, C.~Gonera, S.~Giller and P.~Kosi\'nski,
{\it Galilean invariance in $2+1$ dimensions.}
\texttt{hep-th/9503046} (unpublished).

\bibitem{JaNa}
R.~Jackiw and V.~P.~Nair,
{\it Anyon spin and the exotic central extension of the
planar Galilei group}.
{\sl Phys. Lett.} {\bf B 480} (2000) 237
[\texttt{hep-th/0003130}].
A slight generalization of the JN Ansatz, namely
$s=s_{0}+\kappa c^2$, allows one to have both
non-relativistic spin, $s_{0}$, and exotic structure
$\kappa$, cf. \cite{DHcontr} below.


\bibitem{Dirac}
P. A. M. Dirac,
{\it A positive-energy relativistic wave equation}.
{\sl Proc. R. Soc. Lond. Ser}.
{\bf A322} (1971) 435,  and {\it ibid}.
{\bf A328} (1972) 1.

\bibitem{LL}
J.-M.~L\'evy-Leblond,
{\it Nonrelativistic
Particles and Wave Equations}.
{\sl Comm. Math. Phys.} {\bf 6} (1967) 286.

\bibitem{PPrep}
B. Binegar,
{\it Relativistic field theories in three dimensions}.
  {\sl J. Math. Phys}. {\bf 23} (1982) 1511;
D. R. Grigore,
{\it The projective unitary irreducible representations
of the Poincar\'e group in $2+1$ dimensions}.
{\sl Journ. Math. Phys}. {\bf 34}  (1993) 4172
[\texttt{hep-th/9304142}].

\bibitem{Majorana}
E. Majorana,
{\it Teoria relativistica di particelle con momento
intrinseco arbitrario}.
{\sl Il Nuovo Cimento} {\bf 9} (1932) 335.
The Majorana equation has reemerged in
$2+1$ dimensions in the study of anyons,
see \cite{Plany} and \cite{Ptorsion} below.

\bibitem{Ptorsion}
J. L. Cort\'es and M. S. Plyushchay,
{\it Linear differential equations for a
fractional spin field},
{\sl J. Math. Phys.} {\bf 35} (1994) 6049
[\texttt{hep-th/9405193}].


\bibitem{DHcontr}
C. Duval and P.  A. Horv\'athy,
{\it Spin and exotic Galilean symmetry}
{\sl Phys. Lett}. {\bf B 457} (2002) 306
[\texttt{hep-th/0209166}].

\bibitem{Davyd}
A. S. Davydov,
{\it Quantum Mechanics},  Pergamon Press, New York (1965).

\bibitem{NW}
T. D. Newton, E. P. Wigner,
{\it Localized States for Elementary Systems}.
{\sl Rev. Mod. Phys}. {\bf 21} (1949) 400.


\bibitem{Polyakov}
  A. M. Polyakov,
{\it Fermi-Bose transmutations induced by gauge fields}.
{\sl Mod. Phys. Lett.}  {\bf A3} (1988) 325.

\end{thebibliography}
\end{document}